\documentclass[10pt,twocolumn,journal]{IEEEtran}
\usepackage{tipa}
\usepackage{amsfonts}
\usepackage{bbm}
\usepackage{mathrsfs}
\usepackage{graphicx}
\usepackage{subfigure}
\usepackage{amsmath}
\usepackage{amssymb}
\usepackage{amsbsy}
\usepackage{amstext}
\usepackage{amsopn}
\usepackage{amsthm}
\usepackage{amssymb}
\usepackage{eufrak}
\usepackage{cite}
\usepackage{fancybox}
\usepackage{dashbox}
\usepackage{slashbox}
\usepackage{threeparttable}

\newtheorem{theorem}{Theorem}

\newtheorem{lemma}{Lemma}

\newcommand{\tqbinom}[2]{\genfrac{[}{]}{0pt}{}{#1}{#2}}
\newenvironment{ventry}[1]%
{\begin{list}{}{ 
\settowidth{\labelwidth}{{#1}} 
\setlength{\labelsep}{5mm}
\setlength{\leftmargin}{\labelwidth+\labelsep}}}%
{\end{list}}

\begin{document}
\title{Optimization of Fast-Decodable Full-Rate STBC with Non-Vanishing Determinants}
\author{Tian Peng Ren,~Yong Liang Guan,~Chau Yuen, Yue Zhou and~Er Yang Zhang
\thanks{Manuscript received August 17, 2009; revised February 23 and October 19, 2010. This research is partly supported by the International Design Center (Grant No. G3.2 \& D1.3).}
\thanks{
T. P. Ren and E. Y. Zhang are with the College of Electronic Science and Engineering,
National University of Defense Technology, Changsha 410073, China
(e-mail: tpren@nudt.edu.cn; eyzhang2006@hotmail.com).}
\thanks{
Y. L. Guan is with the School of Electrical and Electronic Engineering, Nanyang Technological University, Singapore 639798 (e-mail: eylguan@ntu.edu.sg).}
\thanks{
C. Yuen is with Singapore University of Technology and Design, Singapore 279623(e-mail: yuenchau@sutd.edu.sg).}
\thanks{
Y. Zhou is with the College of Science, National University of Defense Technology, Changsha 410073, China
(e-mail: gabelozhou@gmail.com).}
\thanks{
Digital Object Identifier ****}
}
\maketitle

\markboth{IEEE Transactions on Communications, Vol. **, No. **, Month ****}
\pubidadjcol

\begin{abstract}
Full-rate STBC (space-time block codes) with non-vanishing determinants achieve the optimal diversity-multiplexing tradeoff but incur high decoding complexity. To permit fast decoding, Sezginer, Sari and Biglieri proposed an STBC structure with special QR decomposition characteristics. In this paper, we adopt a simplified form of this fast-decodable code structure and present a new way to optimize the code analytically. We show that the signal constellation topology (such as QAM, APSK, or PSK) has a critical impact on the existence of non-vanishing determinants of the full-rate STBC. In particular, we show for the first time that, in order for APSK-STBC to achieve non-vanishing determinant, an APSK constellation topology with constellation points lying on square grid and ring radius $\sqrt{m^2+n^2}~(m,n\emph{\emph{ integers}})$ needs to be used. For signal constellations with vanishing determinants, we present a methodology to analytically optimize the full-rate STBC at specific constellation dimension.
\end{abstract}

\IEEEpeerreviewmaketitle

\section{Introduction}
\IEEEPARstart{M}{ulti-input} multi-output (MIMO) systems can be designed to provide two types of gains: transmit diversity gain and spatial multiplexing gain\cite{Zheng}. The full-rate full-diversity space-time block codes (STBC) in \cite{Belfiore,Dayal,Paredes,Sezginer,Rabiei} can achieve both for 2$\times$2 MIMO systems. Recently, a fast-decodable full-rate STBC is proposed by S. Sezginer, H. Sari and E. Biglieri \cite{Sezginer}\cite{Sezginer2}:
\begin{equation}\label{bostbc}
\begin{split}
\textbf{X}_{SSB}=\left[
\begin{array}{cccccccc}
  as_1+bs_3   & -cs^*_2-ds^*_4 \\
  as_2+bs_4   & cs^*_1+ds^*_3
\end{array}
\right]
\end{split}
\end{equation}
where $s_i\in \mathbb{C}$ with $i=1,\cdots,4$ are information symbols, $a,b,c\emph{\emph{ and }}d\in \mathbb{C}$ are design coefficients and $(\cdot)^*$ denotes the complex conjugate. Due to its code structure in (\ref{bostbc}), $\textbf{X}_{SSB}$ has additional zero entries appearing in the upper-triangular matrix after QR decomposition of the equivalent channel matrix, thus making it fast-decodable \cite{Sezginer}\cite{Biglieri}.

It is shown in \cite{Sezginer} that the code structure (\ref{bostbc}) after optimizing for non-vanishing determinant can be rewritten with a single design coefficient. Base on this knowledge, in this paper we adopt a simplified version of the code structure (\ref{bostbc}) by setting $a=1,~b=r,~c=-jr^*\emph{\emph{ and }} d=1$ to obtain:
\begin{equation}\label{rate2coderot}
\begin{split}
\textbf{X}=\left[
\begin{array}{cccccccc}
  s_1+rs_3   & jr^*s^*_2-s^*_4 \\
  s_2+rs_4   & -jr^*s^*_1+s^*_3
\end{array}
\right]
\end{split}
\end{equation}
where $j^2=-1$ and $r \in \mathbb{C}$ is the design coefficient with $\left|r\right|=1$.
Our objective is to analytically optimize the design coefficient $r$ in (\ref{rate2coderot}) to enable the full-rate STBC to achieve non-vanishing determinants. In particular, we will consider the influence of different signal constellation topologies, including rectangular quadrature amplitude modulation (QAM), amplitude-phase shift keying (APSK) and phase shift keying (PSK), on the existence of non-vanishing determinants.
%

The rest of this paper is organized as follows. In Section \ref{secDiversity}, the methods to optimize the design coefficient in (\ref{rate2coderot}) for both integer-coordinate and non-integer-coordinate signal constellations are described. Comparisons of the code in (\ref{rate2coderot}) with other full-rate codes are shown in Section \ref{secSimulation}. This paper is concluded in Section \ref{secConclusion}.

In what follows, bold lower case and upper case letters denote vectors and matrices (sets), respectively; $\mathbb{R}$ and $\mathbb{C}$ denote the real and the complex number fields, respectively; $(\cdot)^R$ and $(\cdot)^I$ stand for the real and imaginary parts of a complex element vector and matrix, respectively; $[\cdot]^H$ denotes the complex conjugate transpose of a matrix; $det(\cdot)$ denotes the determinant of a square matrix.


\section{Optimization of Design Coefficients}\label{secDiversity}
Following \cite{Tarokh}, the diversity gain of $\textbf{X}$ in (\ref{rate2coderot}) is denoted as $rank(\Delta \textbf{X}\cdot \Delta \textbf{X}^H)=rank(\Delta \textbf{X})$ \cite{Beesack}
where the difference matrix $\Delta \textbf{X}=\textbf{X}_1-\textbf{X}_2$, $\textbf{X}_1$ and $\textbf{X}_2$ are STBC matrices based on different information symbols. A full-rank $\Delta \textbf{X}$ guarantees that
$(\Delta \textbf{X}\cdot \Delta \textbf{X}^H)$ is full-rank, and the code $\textbf{X}$ in (\ref{rate2coderot}) will achieve full diversity. When $\Delta \textbf{X}$ is full rank, the coding gain
can be defined as
\begin{equation}\label{codinggain}
\begin{split}
\emph{\emph{Coding gain}}
&\triangleq \underset{\Delta \textbf{X}}{\min}\left[det\left(\Delta \textbf{X}\cdot \Delta\textbf{X}^H\right)\right]\\
&=\underset{\Delta \textbf{X}}{\min}\left(\left| det\left(\Delta\textbf{X}\right)\right|^2 \right)
\end{split}
\end{equation}
where
\begin{equation*}
\begin{split}
det(\Delta \textbf{X})=&-jr^*| \Delta s_1|^2+r| \Delta s_3|^2+\Delta
s_1\Delta s^*_3-j\Delta s^*_1\Delta s_3-\\
&(jr^*| \Delta s_2|^2-r|\Delta s_4|^2-\Delta s_2\Delta s^*_4+j\Delta s^*_2\Delta s_4)\\
=&r(| \Delta s_3|^2+| \Delta s_4|^2)-jr^*(| \Delta s_1|^2+| \Delta
s_2|^2)+\\
&(\Delta s_1\Delta s^*_3+\Delta s_2\Delta s^*_4)-j(\Delta
s^*_1\Delta s_3+\Delta s^*_2\Delta s_4)
\end{split}
\end{equation*}
and $\Delta s_i(i=1,2,3\emph{\emph{ and }}4)$ are the difference
symbols of $s_i$.

\begin{figure}[!t]
\centering
\includegraphics[width=2.7in]{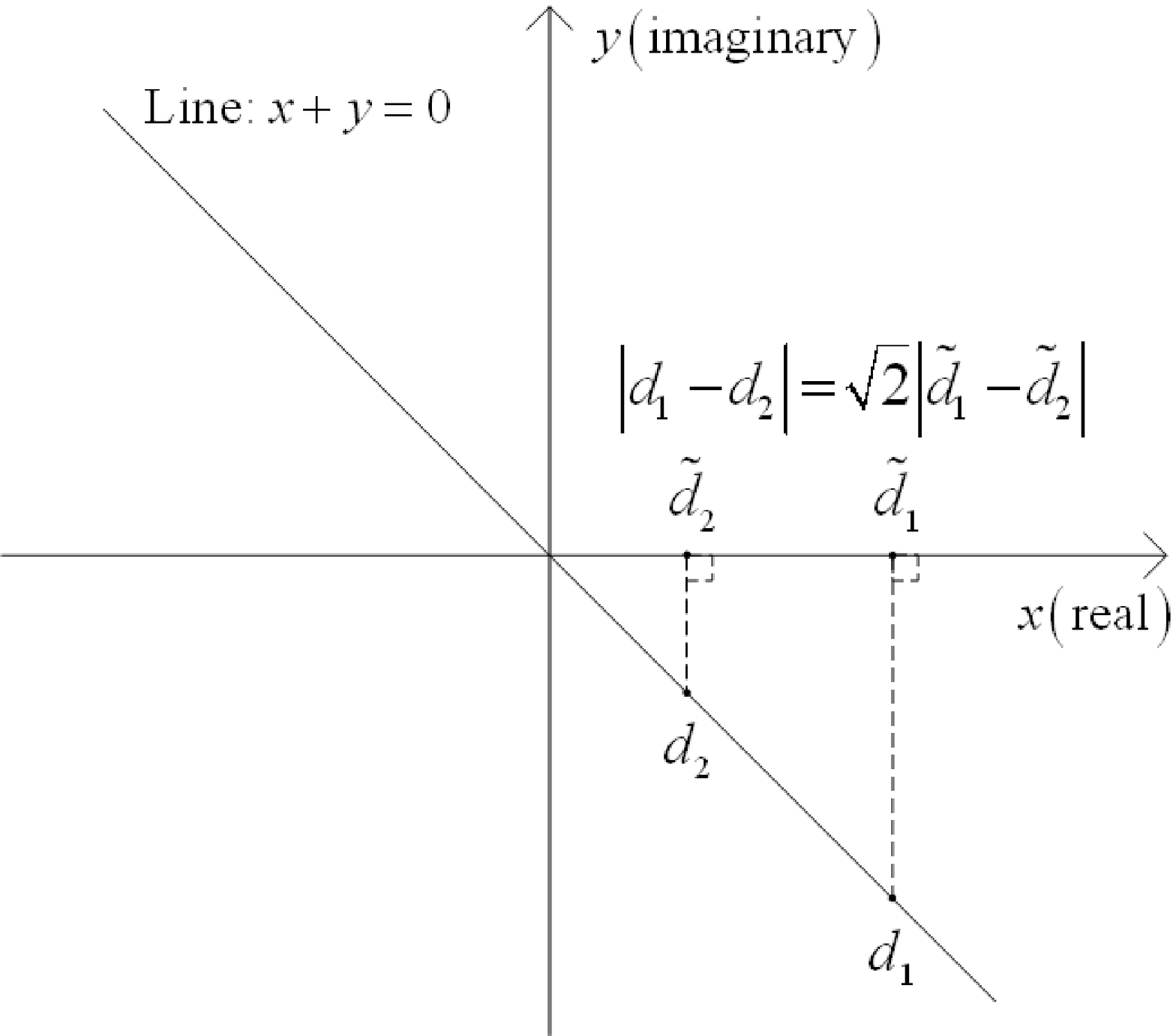}
\caption{$d_1$, $d_2$ and $|d_1-d_2|$ of \textbf{Case I} illustrated in real-imaginary axis graph.} \label{smc2x2_d1d21}
\end{figure}

\begin{figure}
\centering
\includegraphics[width=3.0in]{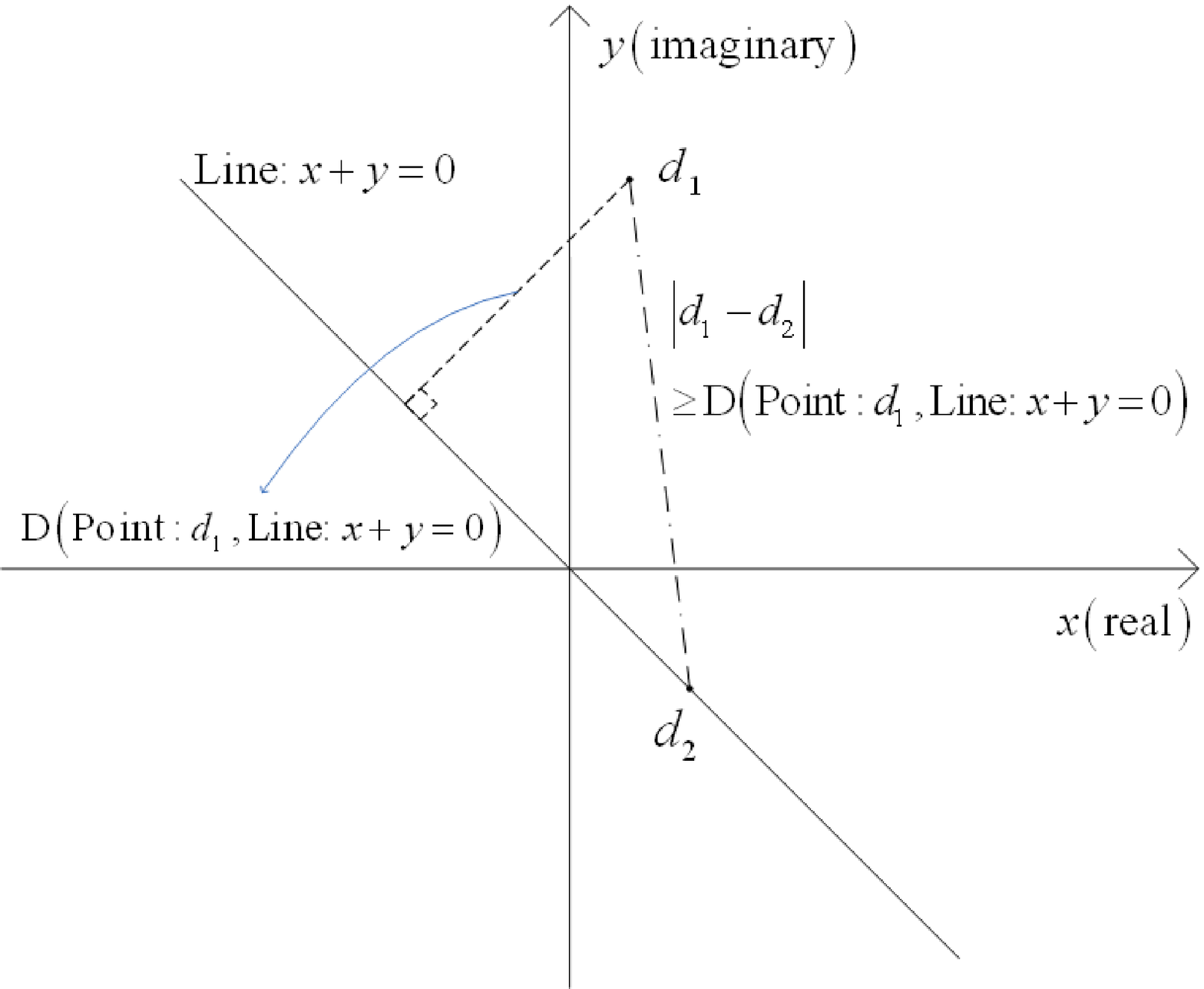}
\caption{$d_1$, $d_2$ and $|d_1-d_2|$ of \textbf{Case II} illustrated in real-imaginary axis graph.} \label{smc2x2_d1d22}
\end{figure}

$det(\Delta \textbf{X})$ can be split into two parts:
\begin{equation}\label{rate2det}
\begin{split}
det(\Delta \textbf{X})&=d_1-d_2
\end{split}
\end{equation}
where
\begin{subequations}
\begin{equation}
~~d_1=r(| \Delta s_3|^2+| \Delta s_4|^2)-jr^*(| \Delta s_1|^2+| \Delta s_2|^2),~~~~~~~~~
\end{equation}
\begin{equation}\label{d_2}
\begin{split} d_2
=&j(\Delta s^*_1\Delta s_3+\Delta s^*_2\Delta s_4)-(\Delta s_1\Delta s^*_3+\Delta s_2\Delta s^*_4)\\
=&[(\Delta s_1\Delta s^*_3+\Delta s_2\Delta s^*_4)^I-(\Delta s_1\Delta s^*_3+\Delta s_2\Delta s^*_4)^R]\\
&(1-j).
\end{split}
\end{equation}
\end{subequations}
Note that $d_1$ is dependent on the design coefficient $r$, while $d_2$ is decided by the difference symbols only. Since $d_2$ is in the form of $(1-j)$ multiplied by a real number (determined by specific values of $\Delta s_1\emph{\emph{ to }}\Delta s_4$), if $d^R_2$ is plotted on the $x$-axis and $d^I_2$ is plotted on the $y$-axis, $d_2$ lies discretely on the line $x+y=0$, as shown in Fig. \ref{smc2x2_d1d21} and Fig. \ref{smc2x2_d1d22}. Since $\underset{\Delta \textbf{X}}{\min}[det(\Delta \textbf{X}\cdot \Delta \textbf{X}^H)]=\underset{\Delta s_1\emph{\emph{ to }}\Delta s_4}{\min}(|d_1-d_2|^2)$, $d_1\neq d_2$ is the necessary and sufficient condition for full diversity, and this can be achieved by influencing $d_1$ using the design coefficient $r$.

Let $r=u+jv$ where $u, v \in \mathbb{R}$ and $u^2+v^2=1$, we have
\begin{equation}  
\begin{split}
d_1
=&r\left(| \Delta s_3|^2+| \Delta s_4|^2\right)-jr^*\left(| \Delta s_1|^2+| \Delta s_2|^2\right)\\
=&\left[\left(| \Delta s_3|^2+| \Delta s_4|^2\right)u-\left(| \Delta s_1|^2+| \Delta s_2|^2\right)v\right]+\\
&\left[\left(| \Delta s_3|^2+| \Delta s_4|^2\right)v-\left(| \Delta s_1|^2+|\Delta s_2|^2\right)u\right]j.
\end{split}
\end{equation}

The coding gain can be analyzed in two different cases as shown below:

\vspace{0.05in}\noindent\underline{\textbf{Case I}: $| \Delta
s_1|^2+| \Delta s_2|^2=| \Delta s_3|^2+| \Delta s_4|^2$}

In this case,
\begin{equation}\label{d_1}
d_1=\left[\left(| \Delta s_1|^2+| \Delta s_2|^2\right)(u-v)\right](1-j).
\end{equation}
Similar to $d_2$, $d_1$ lies on the line $x+y=0$ if
$d^R_1$ is plotted on the $x$-axis and $d^I_1$ is plotted on the
$y$-axis (as illustrated in Fig \ref{smc2x2_d1d21}). The
discrete loci of $d_1$ on the line $x+y=0$ depend not only on
$\Delta s_1\emph{\emph{ to }}\Delta s_4$, but also on $u$ and $v$
(the design coefficients).

Let $d_1=\tilde d_1(1-j)$ and $d_2=\tilde d_2(1-j)$, from (\ref{d_1}) and (\ref{d_2}) we get
\begin{subequations}
\begin{align}
\tilde d_1=&\left(| \Delta s_1|^2+| \Delta
s_2|^2\right)(u-v)
\\
\tilde d_2=&\left(\Delta s_1\Delta s^*_3+\Delta s_2\Delta
s^*_4\right)^I-\left(\Delta s_1\Delta s^*_3+\Delta s_2\Delta
s^*_4\right)^R
\end{align}
\end{subequations}

To achieve full diversity gain ($d_1\neq d_2$), $u$ and $v$ must be chosen to achieve $\tilde d_1\neq
\tilde d_2$. Note that 
$(u-v)\in [-\sqrt{2},~\sqrt{2}]$ due to $u^2+v^2=1$. Hence, in this case the \textbf{Case I} coding gain is
\begin{equation} \label{rate2det2}
\begin{split}
 &\underset{\Delta \textbf{X}}{\min}\left[det\left(\Delta \textbf{X}\cdot \Delta \textbf{X}^H \right) \right]\\
=&\underset{\Delta s_1\emph{\emph{ to }}\Delta s_4}{\min}\left(|d_1-d_2|^2 \right)\\
=&\underset{\Delta s_1\emph{\emph{ to }}\Delta s_4}{\min}\left(2| \tilde d_1-\tilde d_2|^2\right).
\end{split}
\end{equation}

\vspace{0.05in} \noindent \underline{\textbf{Case II}: $| \Delta
s_1|^2+| \Delta s_2|^2\neq | \Delta s_3|^2+| \Delta s_4|^2$}

In this case, we have
\begin{equation}
\begin{split}
 &d^R_1+d^I_1\\
=&\left(| \Delta s_3|^2+| \Delta s_4|^2\right)u-\left(| \Delta s_1|^2+|\Delta s_2|^2\right)v+\\
&\left(| \Delta s_3|^2+| \Delta s_4|^2\right)v-\left(| \Delta s_1|^2+| \Delta s_2|^2\right)u\\
=&\left(| \Delta s_3|^2+| \Delta s_4|^2-| \Delta s_1|^2-| \Delta
s_2|^2\right)(u+v)\\
=&\emph{\emph{ a non-zero real number }}\times (u+v)
\end{split}
\end{equation}

If $u+v\neq 0$, then $d^R_1+d^I_1\neq 0$ and $d_1$ will never lie on the line $x+y=0$, as shown in Fig. \ref{smc2x2_d1d22}. Since we have earlier shown that $d_2$ always lies on the line $x+y=0$, it implies that $d_1\neq d_2$, hence full diversity is always achieved by \textbf{Case II}. As shown in Fig. 2, the Euclidean distance between $d_1$ and $d_2$ can be lower bounded by the perpendicular distance between $d_1$ and the line $x+y=0$ where $d_2$ lies on. \ref{smc2x2_d1d22}. Hence, the \textbf{Case II} coding gain is lower bounded as
\begin{equation}\label{rate2det1}
\begin{split}
&\underset{\Delta \textbf{X}}{\min}\left[det\left(\Delta \textbf{X}\cdot \Delta \textbf{X}^H \right) \right]\\
=&\underset{\Delta s_1\emph{\emph{ to }}\Delta s_4}{\min}\left(|d_1-d_2|^2 \right)\\
\geq&\underset{\Delta s_1\emph{\emph{ to }}\Delta s_4}{\min}\left[ \emph{\emph{D}}^2(\emph{\emph{Point}}:
d_1,~\emph{\emph{Line}}: x+y=0)\right]
\\=&\underset{\Delta s_1\emph{\emph{ to }}\Delta s_4}{\min}\left[(| \Delta s_3|^2+| \Delta s_4|^2-| \Delta s_1|^2-| \Delta s_2|^2)^2\right.\\
&\left.(u+v)^2/2\right]
\end{split}
\end{equation}
where $\emph{\emph{D}}(\emph{\emph{Point, Line}})$ denotes the perpendicular distance from a point to a line.

Next, we will find the $u$ and $v$ that satisfy the above full diversity conditions and maximize the coding gain
$\underset{\Delta \textbf{X}}{\min}[det(\Delta \textbf{X}\cdot \Delta \textbf{X}^H)]$ for the above two cases. To make the optimization process tractable, we will first find the $u$ and $v$ that maximize the \textbf{Case I} coding gain (\ref{rate2det2}). Then we show that the \textbf{Case II} coding gain (\ref{rate2det1}) substituted with the $u$ and $v$ obtained are larger than the maximized (\ref{rate2det2}). Hence the \textbf{Case I} coding gain (\ref{rate2det2}) dominates the performance of the code $\textbf{X}$, and the $u$ and $v$ obtained by maximizing (\ref{rate2det2}) will be the global optimum design coefficients.

\subsection{Integer-Coordinate Signal Constellations}
When integer-coordinate signal constellations (such as rectangular QAM) are applied, the difference symbols also have integer coordinates \cite{Belfiore}\cite{Biglieri}, where the minimum Euclidean distance in the signal constellation is fixed at 1.
\begin{lemma}\label{lemma3_conclusion}
\emph{When integer-coordinate signal constellations are applied, the coding gain (\ref{rate2det2}) of the code $\textbf{X}$ in \textbf{Case I} is upper bounded by $1/2$, and the maximum value can be achieved if and only if
$u-v=\pm1/2$.}
\end{lemma}
\begin{IEEEproof}
In Appendix \ref{proof_theorem1}.
\end{IEEEproof}

The following theorem establishes the non-vanishing determinant of $\Delta \textbf{X}\cdot \Delta \textbf{X}^H$ with
integer-coordinate signal constellations.

\begin{theorem} \label{theorem1}
\emph{When integer-coordinate signal constellations are applied, the optimum design coefficient $r=u+jv$ to achieve full diversity and maximum non-vanishing coding gain for the code $\textbf{X}$ in (\ref{rate2coderot}) is given by:}
\begin{subequations} \label{rate2det_con11}
\begin{align}
&u=\left(1\pm\sqrt{7}\right)/4,~~~v=\left(-1\pm\sqrt{7}\right)/4\\
\emph{\emph{\emph{or}}}~~~~&u=\left(-1\pm\sqrt{7}\right)/4,~v=\left(1\pm\sqrt{7}\right)/4.
\end{align}
\end{subequations}
\end{theorem}
\begin{IEEEproof}
Let us first consider \textbf{Case I}. (\ref{rate2det_con11}) can be obtained by combining $u-v=\pm1/2$ from Lemma
\ref{lemma3_conclusion} and $u^2+v^2=1$ (by definition).

Next, for \textbf{Case II}, since $| \Delta s_1|^2+| \Delta
s_2|^2\neq | \Delta s_3|^2+| \Delta s_4|^2$ and integer-coordinate
signal constellations are applied, we have $ \left|| \Delta s_3|^2+|
\Delta s_4|^2-| \Delta s_1|^2-| \Delta s_2|^2\right| \geq1$.
Substituting the $u$ and $v$ in (\ref{rate2det_con11}) to
(\ref{rate2det1}), the coding gain becomes
\begin{equation}\label{redet2}
\begin{split}
&\underset{\Delta \textbf{X}}{\min}\left[det\left(\Delta \textbf{X}\cdot \Delta \textbf{X}^H \right) \right]\\
\geq&\underset{\Delta s_1\emph{\emph{ to }}\Delta s_4}{\min} \left[(| \Delta s_3|^2+| \Delta s_4|^2-| \Delta s_1|^2-| \Delta s_2|^2)^2\right.\\
&\left.(u+v)^2/2\right]\\
=&7/8.
\end{split}
\end{equation}

Comparing the \textbf{Case II} coding gain expressed in (\ref{redet2}) with the \textbf{Case I} coding gain expressed in Lemma \ref{lemma3_conclusion}, the \textbf{Case I} coding gain is lower, hence it is the overall coding gain, and Theorem \ref{theorem1} is proved.
\end{IEEEproof}

\vspace{0.02in}
\emph{Remark}: The method of proof in this paper, specifically \textbf{Case I} and \textbf{Case II}, are presented in a different way than the proof provided in \cite{Sezginer}. Interestingly, however, the optimized design coefficients in both papers are found to be the same.

\vspace{0.07in}
\noindent \underline{\textbf{Application 1}: Integer-coordinate APSK}

APSK (amplitude-phase shift keying) is a high-order modulation scheme commonly used in SISO (single-input single-output) communications. Conventional APSK topology resembles multi-ring PSK, or circular QAM, as illustrated in Fig. \ref{fig_8apsk} and Fig. \ref{fig_16apsk}. Compared with rectangular QAM,
APSK has advantages such as lower constellation peak-to-average-power ratio (PAPR) and robustness against nonlinear distortion in SISO communications \cite{Miller}. Moreover, APSK may lead to larger minimum Euclidean distance per unit average power for certain constellation dimension such as 8-APSK \cite{Proakis}. Hence APSK has been adopted by the DVB-S2 Standard \cite{dvb-s2}.

In order for the APSK constellation with arbitrary constellation dimension to achieve non-vanishing coding gain with the code $\textbf{X}$ in (\ref{rate2coderot}), we may deduce from Theorem \ref{theorem1} that:

(1) The APSK constellation points should lie on square grids and ring radius $\sqrt{m^2+n^2}~(m,n\emph{\emph{ integers}})$;

(2) The design coefficient in (\ref{rate2det_con11}) should be adopted for the code $\textbf{X}$.

Two examples of the proposed APSK constellations are shown in Fig. \ref{fig_8apskm} and Fig. \ref{fig_16apskm}. With minimum Euclidean distance fixed at 1, they lead to a non-vanishing coding gain of $1/2$ for the code $\textbf{X}$ (same proof as Theorem \ref{theorem1}).

\subsection{Non-Integer-Coordinate Signal Constellations} \label{section_II_non_integer}
When non-integer-coordinate signal constellations such as $M$-ary phase shift keying ($M$-PSK) are applied, the difference symbols $\Delta s_i(i=1,2,3$ and $4)$ do not have integer coordinates. This leads to a vanishing determinant for the code $\textbf{X}$ in (\ref{rate2coderot}) even when the minimum Euclidean distance is fixed at 1. The proof is straightforward, hence omitted.

Although the code $\textbf{X}$ in (\ref{rate2coderot}) with $M$-PSK constellations has vanishing determinant, the code can still be analytically optimized for a specific constellation size based on the mathematical framework presented earlier. The optimization methodology is described below:
\begin{ventry}{Step 2}
\item[Step 1] Consider \textbf{Case I}: $| \Delta s_1|^2+| \Delta s_2|^2= | \Delta s_3|^2+| \Delta s_4|^2$, whose coding gain expression is shown in (\ref{rate2det2}). Given a signal constellation, find out all the values of $\left(| \Delta s_1|^2+| \Delta s_2|^2\right)$; For each value of $\left(| \Delta s_1|^2+| \Delta s_2|^2\right)$, find out all the values of $\tilde d_2$. Since $ \tilde d_1$ is a function of $(u-v)$, the expression of $|\tilde d_1-\tilde d_2|$ as a function of $(u-v)$ can be evaluated. 
    Based on these expressions of $|\tilde d_1-\tilde d_2|$ and $(u-v)\in[-\sqrt{2},~\sqrt{2}]$, obtain the maximum value of $\underset{\Delta s_1\emph{\emph{ to }}\Delta s_4}{\min}|\tilde d_1-\tilde d_2|$, and the corresponding $(u-v)$. Combining the $(u-v)$ obtained with $u^2+v^2=1$, we can obtain the corresponding $u$, $v$ and the maximized coding gain;
\item[Step 2] Next, consider \textbf{Case II}: $| \Delta s_1|^2+| \Delta s_2|^2\neq | \Delta s_3|^2+| \Delta s_4|^2$. Substitute the $u$ and $v$ obtained in Step 1 into (\ref{rate2det1}) to obtain the \textbf{Case II} coding gain. If the \textbf{Case II} coding gain is higher than that of \textbf{Case I}, then the latter is the overall coding gain by definition, and we conclude that the code $\textbf{X}$ in (\ref{rate2coderot}) with design coefficients $r=u+jv$ obtained in \textbf{Case I} achieves full diversity gain and maximum coding gain. For PSK and conventional APSK, this is found to be always true.
\end{ventry}

\vspace{0.07in}
\noindent \underline{\textbf{Application 2}: 8-PSK}

Applying the optimization steps described above to the code $\textbf{X}$ in (\ref{rate2coderot}) with 8-PSK constellation, the optimum design coefficients shown in (\ref{rate2det_con21}) and the maximum coding gain of $(22572-15912\sqrt{2})/2401$ are obtained.
\begin{subequations}\label{rate2det_con21}
\begin{align}
&u=\left( 11+6\sqrt{2}\pm \sqrt{4609-132\sqrt{2}}\right)/98,\\
&v=\left(-11-6\sqrt{2}\pm \sqrt{4609-132\sqrt{2}}\right)/98\\
\emph{\emph{or}}~~~~~
&u=\left(-11-6\sqrt{2}\pm \sqrt{4609-132\sqrt{2}}\right)/98,\\
&v=\left(11+6\sqrt{2}\pm \sqrt{4609-132\sqrt{2}}\right)/98.
\end{align}
\end{subequations}

\vspace{0.07in}
\noindent \underline{\textbf{Application 3}: Conventional APSK}

\begin{figure}
\centering
\subfigure [Conventional 8-QAM \cite{xia_rotation} with $a=\frac{1}{\sqrt{78}}$]
  {
    \label{fig_8qam} 
    \includegraphics[width=1.5in]{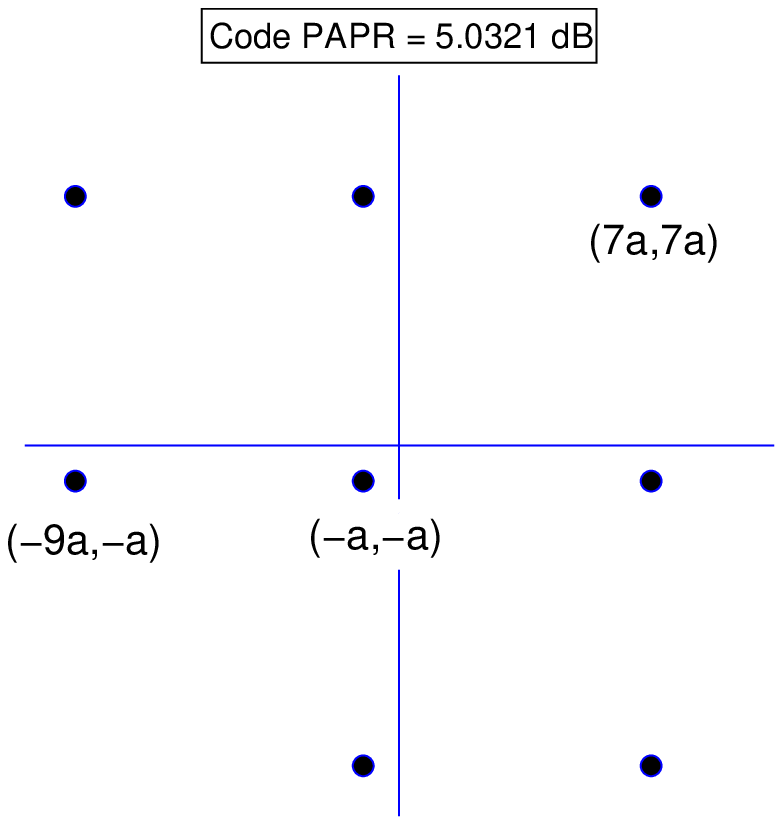}
    }
\hspace{0.1in}
\subfigure [Conventional 8-APSK \cite{Proakis} with $b=\frac{1}{\sqrt{3+\sqrt{3}}}$]
  {
    \label{fig_8apsk} 
    \includegraphics[width=1.52in]{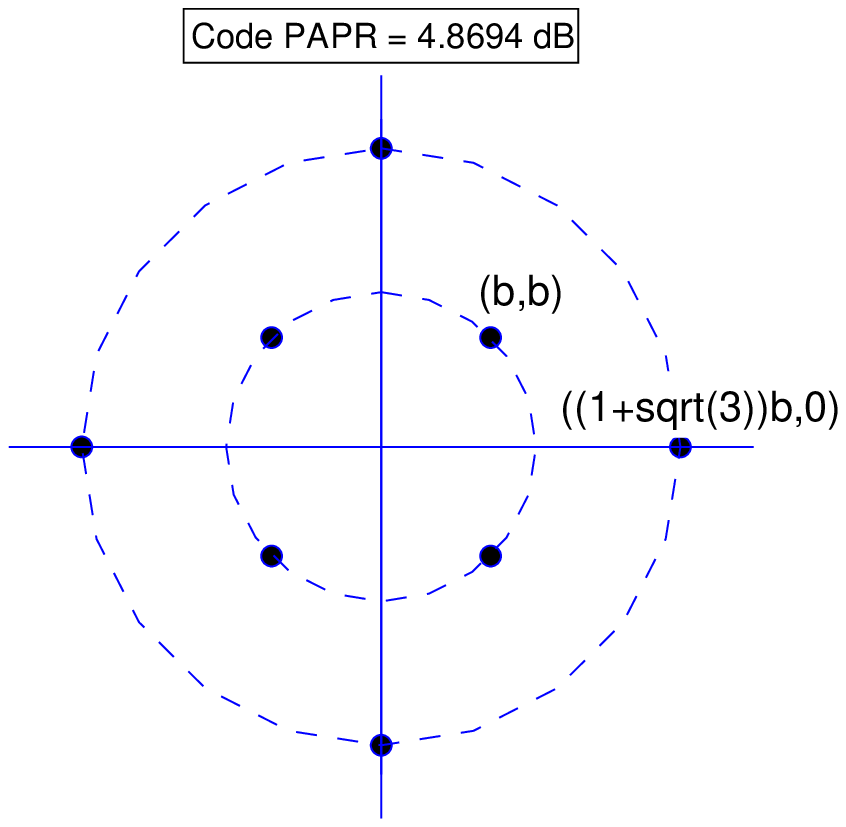}
    }
\hspace{0.1in}
\subfigure[Proposed 8-APSK with $c=\frac{1}{\sqrt{3}}$]
  {
    \label{fig_8apskm} 
    \includegraphics[width=1.5in]{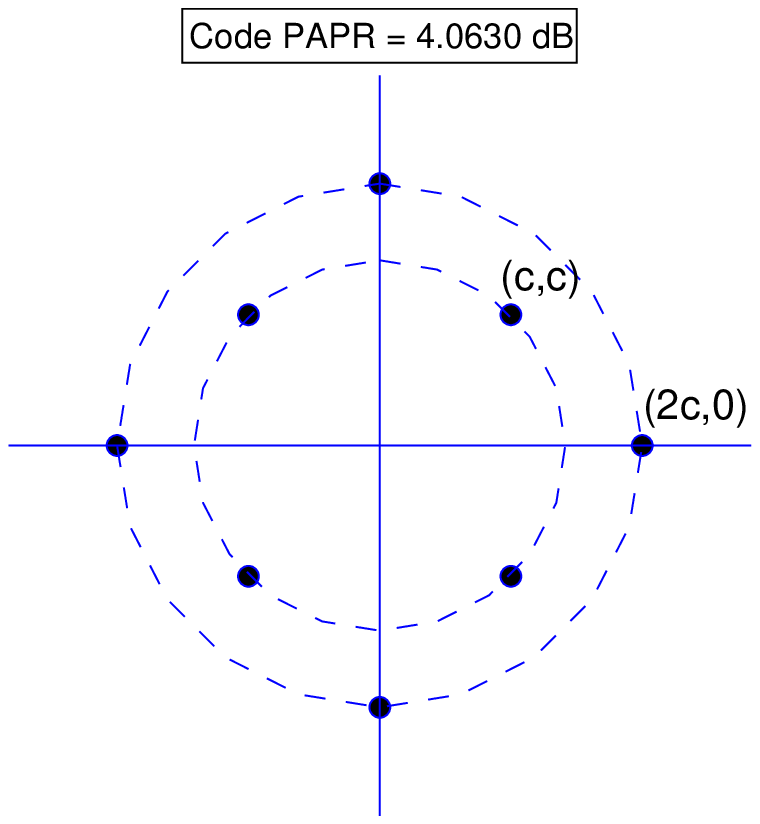}
    }
\hspace{0.1in}
\subfigure [Conventional 16-QAM with $d=\frac{1}{\sqrt{10}}$]
   {
    \label{fig_16qam} 
    \includegraphics[width=1.5in]{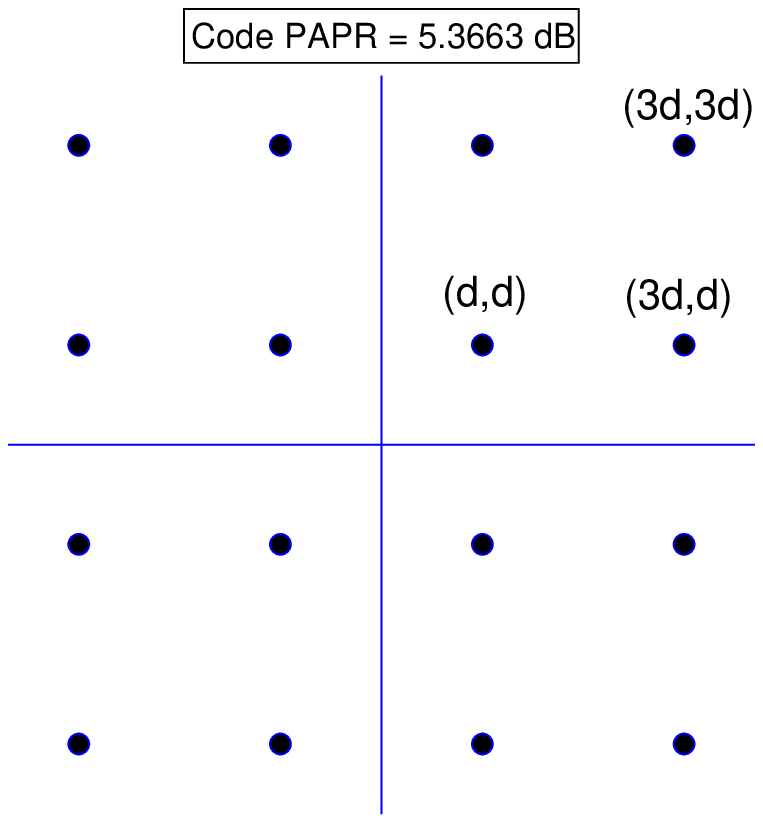}
    }
\hspace{0.1in}
\subfigure [Conventional 16-APSK \cite{dvb-s2} with $r_1=\frac{2}{\sqrt{13+6\sqrt{3}}}$ and $r_2=\frac{2\sqrt{2}}{\sqrt{8-\sqrt{3}}}$]
   {
    \label{fig_16apsk} 
    \includegraphics[width=1.5in]{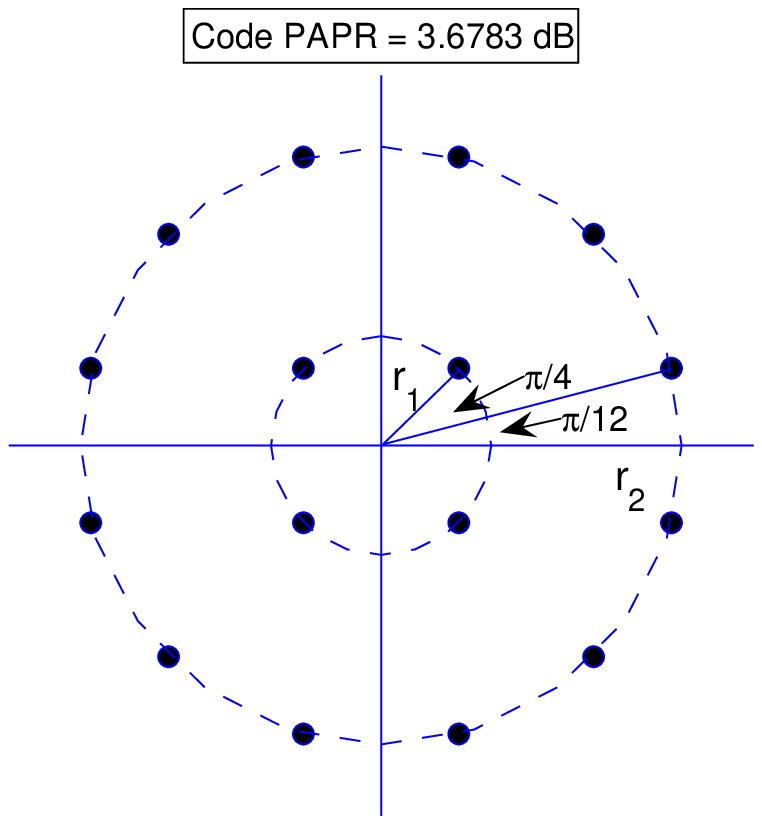}
    }
\hspace{0.1in}
\subfigure [Proposed 16-APSK with $f=0.5$]
  {
    \label{fig_16apskm} 
    \includegraphics[width=1.5in]{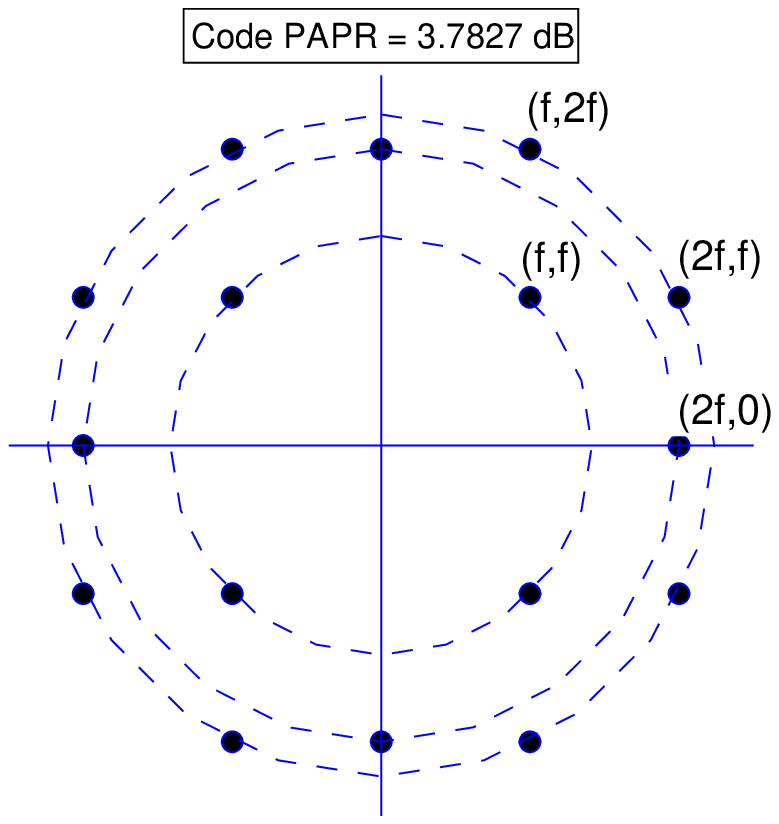}
    }
  \caption{QAM, conventional APSK and the proposed APSK with unit-average-power symbols (sqrt($\cdot$) means $\sqrt{\cdot}$).} \label{fig_cQAM}
\end{figure}

Similarly, the optimized design coefficients and coding gains for the conventional 8-APSK shown in Fig. \ref{fig_8apsk} and the conventional 16-APSK shown in Fig. \ref{fig_16apsk} can be found, and are listed in Table \ref{table_apsk}. They will be used later in Fig. \ref{8apsk}.

\section{Simulations and Discussions} \label{secSimulation}
In the simulations, we assume that the Rayleigh fading channel is quasi-static in the sense that the channel coefficients do not change within a codeword, and the channel state information (CSI) is perfectly known at the receiver.

\subsection{Integer-Coordinate Signal Constellations}
\begin{figure}[!t]
\centering
\includegraphics[width=3.7in]{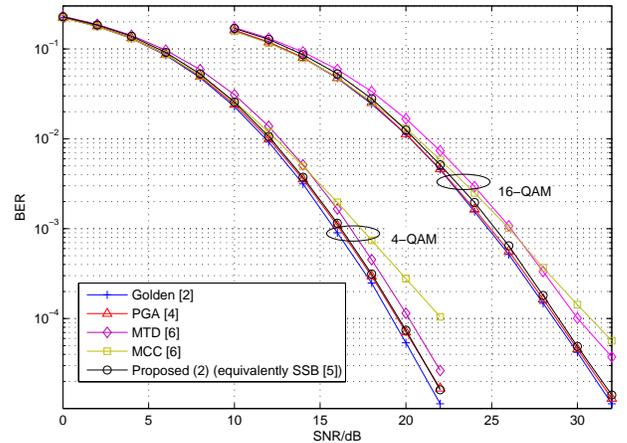}
\caption{ML decoding performances of different full-rate codes in 2$\times$2 MIMO systems with 4-QAM and 16-QAM constellations.} \label{fig_qr21}
\end{figure}

Firstly, we show the ML bit error rate (BER) performances of the proposed code $\textbf{X}$ in (\ref{rate2coderot}) with the optimized coefficient (\ref{rate2det_con11}), the Golden \cite{Belfiore}, PGA \cite{Paredes}, MTD and MCC \cite{Rabiei} codes\footnote{PGA, MTD and MCC denote the Paredes-Gershman-Alkhanari, Maximum Transmit Diversity and Maximum Channel Capacity codes, respectively.} for 2$\times$2 MIMO systems with 4-QAM and 16-QAM in Fig. \ref{fig_qr21}. The SSB code \cite{Sezginer} is equivalent to the proposed code $\textbf{X}$. The results show that the proposed code $\textbf{X}$ in (\ref{rate2coderot}) with design coefficients (\ref{rate2det_con11}) has BER performance slightly worse than Golden code \cite{Belfiore}, comparable with PGA code \cite{Paredes}, and better than MTD and MCC codes \cite{Rabiei}.

On the other hand, as the proposed code structure (\ref{rate2coderot}) is fast-decodable, it has computational complexity order $M^2$ \cite{Sezginer}, same as the codes in \cite{Paredes}\cite{Rabiei}. Since the computational complexity order of Golden code \cite{Belfiore} is $M^4$, the small performance loss of the proposed code compared to Golden code can be viewed as a small penalty to be paid for the complexity reduction.

\subsection{Non-Integer-Coordinate Signal Constellations}
\begin{figure}[!t]
\centering
\includegraphics[width=3.83in]{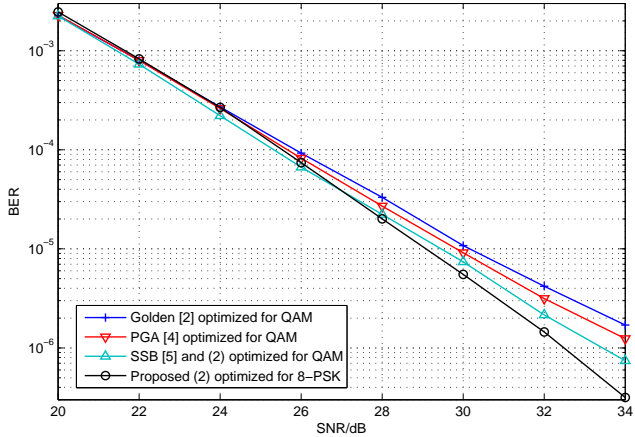}
\caption{ML decoding performances of different full-rate codes in 2$\times$2 MIMO systems with 8-PSK constellation.} \label{fig_qr22}
\end{figure}

The BER performance of the code $\textbf{X}$ in (\ref{rate2coderot}) is next compared with other full-rate codes
\cite{Belfiore}\cite{Paredes}\cite{Sezginer} with 8-PSK in Fig. \ref{fig_qr22}. Here the design coefficients in (\ref{rate2det_con21}) are adopted for the code \textbf{X} in (\ref{rate2coderot}), while the optimum design coefficients for the other codes are taken from their respective publications. From the simulation results, we can see that the code \textbf{X} in (\ref{rate2coderot}) achieves a larger BER slope when the SNR is high. This is because the other codes, including Golden code, were optimized for QAM, not PSK.

The coding gains of the full-rate STBC\textquoteright s with QAM and PSK constellations are tabulated in Table \ref{table_codinggain} with the average power of information symbols normalized to 1. In all cases, they concur with the BER observations made in Fig. \ref{fig_qr21} and \ref{fig_qr22}.

\begin{center}
\begin{threeparttable}[!b]
\tabcolsep 0.8mm
\caption{Coding Gain Comparisons (Considering Unit-Average-Power Information Symbols).} \label{table_codinggain}
\newcommand{\rb}[1]{\raisebox{1.2ex}[0pt]{#1}}
\newcommand{\rbb}[1]{\raisebox{1.0ex}[0pt]{#1}}
{\small\begin{tabular}{|c||c|c|c|} 
\hline 2$\times$2 STBC & 4-QAM& 16-QAM & 8-PSK$^a$
\\\hline\hline
Golden \cite{Belfiore}& 3.2 & 0.128 & $\backslash$
\\ \hline
PGA \cite{Paredes} & 2.286 & 0.0914 & $\backslash$
\\ \hline
SSB \cite{Sezginer} &2 & 0.08 & $\backslash$
\\ \hline
MTD \cite{Rabiei} &0.64 & 0.0022 & $\backslash$
\\ \hline
MCC \cite{Rabiei} &Non-full diversity & Non-full diversity & $\backslash$
\\ \hline
Proposed code &  & &
\\
\textbf{X} in (\ref{rate2coderot}) & \rb{2} & \rb{0.08} & \rb{0.0288}
\\ \hline
\end{tabular}}
\footnotesize{
\begin{tablenotes}
\item[$^a$] In \cite{Belfiore,Dayal,Paredes,Sezginer,Rabiei}, the code design coefficients for 8-PSK are not given.
\end{tablenotes}}
\end{threeparttable}
\end{center}

\subsection{APSK Constellations}
\begin{figure}[!t]
\centering
\includegraphics[width=3.7in]{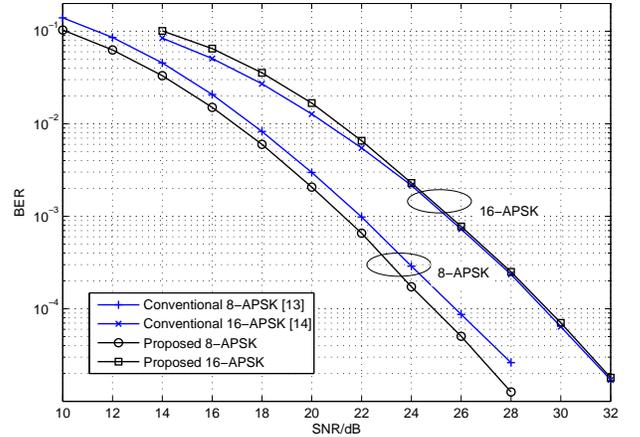}
\caption{ML decoding performances of the code $\textbf{X}$ in (\ref{rate2coderot}) in 2$\times$2 MIMO systems with the conventional and proposed 8/16-APSK constellations shown in Fig. \ref{fig_cQAM}.} \label{8apsk}
\end{figure}

Comparisons of the properties and performance of the code $\textbf{X}$ in (\ref{rate2coderot}), when used with the conventional APSK topology versus the proposed APSK topology shown in Fig. \ref{fig_cQAM}, are presented in Table \ref{table_apsk} and Fig. \ref{8apsk}, respectively. Note that Fig. \ref{fig_8apsk} is the best known conventional 8-APSK (in SISO sense), while Fig. \ref{fig_16apsk} is the 16-APSK adopted by the DVB-S2 Standard \cite{dvb-s2}. In the BER simulations, the corresponding optimum code design coefficients $r$\textquoteright s from Table \ref{table_apsk} are applied. Note from Table \ref{table_apsk} that the proposed APSK does not need to change its design coefficient $r$ for different constellation dimensions as it achieves non-vanishing determinant, but this is not true for the conventional APSK.

Interestingly, Table \ref{table_apsk} shows that although the proposed APSK shown in Fig. \ref{fig_8apskm} and Fig. \ref{fig_16apskm} have smaller minimum Euclidean distance (hence lower PAPR for the proposed 8-APSK), they achieve higher coding gain than the conventional APSK. This is because the coding gains do not depend linearly nor solely on the minimum Euclidean distance, as shown in (\ref{rate2det2}) and (\ref{rate2det1}).

Fig. \ref{8apsk} shows that the code $\textbf{X}$ in (\ref{rate2coderot}) with the proposed 8-APSK has much better performance than the conventional 8-APSK, while the proposed 16-APSK has similar performance as the conventional 16-APSK at high SNR. Fig. \ref{8apsk} also testifies that the code design coefficients shown in Table \ref{table_apsk} for the conventional 8/16 APSK achieve full diversity.
\begin{center}
\begin{threeparttable}[!b]
\tabcolsep 0.8mm
\caption{Comparisons of Conventional and Proposed APSK for the Code \textbf{X} in (\ref{rate2coderot}) (Considering
Unit-Average-Power Information Symbols).} \label{table_apsk}
\newcommand{\rb}[1]{\raisebox{1.2ex}[0pt]{#1}}
\newcommand{\rbb}[1]{\raisebox{1.0ex}[0pt]{#1}}
{\small\begin{tabular}{|c||c|c|c|c|c|}  
\hline
{\rule[-0mm]{0mm}{1mm}}          & Minimum             &   Code design        &
\\
{\rule[-0.5mm]{0mm}{1mm}} \rb{APSK} &  Euclidean dis. &    coefficient$^b$: $r$ & \rb{Coding gain}
\\
\hline \hline
Conventional &   &     &
\\
8-APSK \cite{Proakis}  & \rb{0.9194} &  \rb{$0.9454 + j0.3258$}  & \rb{0.0230}
\\ \hline
Proposed                &   &    &
\\
8-APSK                     & \rb{0.8165} &  \rb{$0.9114 + j0.4114$}  & \rb{0.2222}
\\ \hline
Conventional  &   &    &
\\
16-APSK \cite{dvb-s2}  & \rb{0.5848} &  \rb{$0.8294 + j0.5587$}  & \rb{0.0004}
\\ \hline
Proposed           &     &    &
\\
16-APSK            & \rb{0.5}    &  \rb{$0.9114 + j0.4114$}  & \rb{0.03125}
\\ \hline
\end{tabular}}
\footnotesize{
\begin{tablenotes}
\item[$^b$] For conventional APSK, the code design coefficient $r$ is optimized following the optimization methodology shown in Section \ref{section_II_non_integer}; For the proposed APSK, the code design coefficient $r=(1+\sqrt{7})/4+j(-1+\sqrt{7})/4\approx 0.9114 + j0.4114$ from (\ref{rate2det_con11}) is used.
\end{tablenotes}}
\end{threeparttable}
\end{center}

\section{Conclusions}\label{secConclusion}
In this paper, a one-parameter full-rate STBC code structure with fast ML decoding capability adapted from \cite{Sezginer}, $\textbf{X} = \tqbinom{s_1+rs_3~~~jr^*s^*_2-s^*_4}{s_2+rs_4~-jr^*s^*_1+s^*_3}$, is analyzed for non-vanishing determinant. When used with integer-coordinate signal constellations such as rectangular QAM, the code design coefficient $r$ is analytically optimized to achieve maximum non-vanishing determinants. When used with non-integer coordinate constellations such as $M$-PSK, the STBC is found to have vanishing determinants even when the minimum Euclidean distance is fixed at 1. For such vanishing-determinant cases, an analytical methodology is presented to optimize the code to achieve maximum coding gain for a specific constellation dimension. In addition, we consider for the first time the use of APSK constellations in the fast-decodable full-rate STBC, and we show that the APSK-STBC can achieve non-vanishing determinant not with the conventional APSK topologies, but with a APSK topology with constellation points lying on square grid and ring radius $\sqrt{m^2+n^2}~(m,n\emph{\emph{ integers}})$. The corresponding optimum STBC design coefficient $r$ and non-vanishing coding gain are $\left(1\pm\sqrt{7}\right)/4+j\left(-1\pm\sqrt{7}\right)/4$ (or $\left(-1\pm\sqrt{7}\right)/4+j\left(1\pm\sqrt{7}\right)/4$) and $1/2$, respectively. BER simulation, coding gain and code PAPR (peak to average power ratio) enumeration results show that the proposed APSK topology leads to lower code PAPR than QAM, and better or similar BER at high SNR.

\appendix  \label{proof_theorem1}

Let us first introduce Lemma \ref{lemma1_conclusion} and Lemma \ref{lemma2_conclusion} which will be used later to prove Lemma \ref{lemma2_conclusion} and Lemma \ref{lemma3_conclusion}, respectively. In the following, $b|a$ denotes that $b$ divides $a$, and $b\nmid a$ denotes that $b$ cannot divide $a$.

\begin{lemma}\label{lemma1_conclusion}
\emph{For integers $a,b,c,d\emph{\emph{ and }}k$, if
$2^{2k}|a^2+b^2+c^2+d^2$, then}
\begin{equation*}
\begin{split}
\emph{either~~~~~~}
&2^{k-1}|a,~2^{k-1}|b, ~2^{k-1}|c, ~2^{k-1}|d\emph{ and }\\
&2^{k}|a, ~2^{k}|b, ~2^{k}|c,~2^{k}|d,\\
\emph{or~~~~~~~~~~}
& 2^{k-1}|a, ~2^{k-1}|b, ~2^{k-1}|c, ~2^{k-1}|d\emph{ and }\\
&2^{k}\nmid a, ~2^{k}\nmid b, ~2^{k}\nmid c, ~2^{k}\nmid d
\end{split}
\end{equation*}
\emph{will hold.}
\end{lemma}
\begin{IEEEproof}
The proof is provided by induction on $k$.

\underline{Case $k=1$}: Clearly, $1|a$, $1|b$, $1|c$ and $1|d$.

As the value of $a^2\emph{\emph{ mod }}4$ is equal to 0 or 1 for any
integer $a$ and $4|a^2+b^2+c^2+d^2$, one of the following two
equations must hold
\begin{equation*}
a^2\emph{\emph{ mod }}4=b^2\emph{\emph{ mod }}4=c^2\emph{\emph{ mod }}4=d^2\emph{\emph{ mod }}4=0,
\end{equation*}
\begin{equation*}
a^2\emph{\emph{ mod }}4=b^2\emph{\emph{ mod }}4=c^2\emph{\emph{ mod }}4=d^2\emph{\emph{ mod }}4=1.
\end{equation*}
In other words, $2|a$, $2|b$, $2|c$, $2|d$ must hold at the same
time, or $2\nmid a$, $2\nmid b$, $2\nmid c$, $2\nmid d$ must hold at
the same time.

\underline{Case $k=2$}: Since $16|a^2+b^2+c^2+d^2$, we have
$8|a^2+b^2+c^2+d^2$. As the value of $a^2\emph{\emph{ mod }}8$ is
equal to 1 for any odd integer $a$, it follows from
$8|a^2+b^2+c^2+d^2$ that $a,b,c\emph{\emph{ and }}d$ are even
integers, i.e., $2^{2-1}|a$, $2^{2-1}|b$, $2^{2-1}|c$ and
$2^{2-1}|d$.

Then, $4|(\frac{a}{2})^2+(\frac{b}{2})^2+(\frac{c}{2})^2+(\frac{d}{2})^2$
where $\frac{a}{2}$, $\frac{b}{2}$, $\frac{c}{2}$ and $\frac{d}{2}$
are integers. Applying the conclusions in Case $k=1$, we have
$2^{2}|a$, $2^{2}|b$, $2^{2}|c$, $2^{2}|d$ at the same time, or
$2^{2}\nmid a$, $2^{2}\nmid b$, $2^{2}\nmid c$, $2^{2}\nmid d$ at
the same time.

\underline{Case $k>2$}: Now Let $k-1$ be the induction hypothesis, we
prove the induction step.

Since $2^{2k}|a^2+b^2+c^2+d^2$, $2^{k-2}|a$, $2^{k-2}|b$,
$2^{k-2}|c$ and $2^{k-2}|d$, we have
$16|(\frac{a}{2^{k-2}})^2+(\frac{b}{2^{k-2}})^2+(\frac{c}{2^{k-2}})^2+(\frac{d}{2^{k-2}})^2$
where $\frac{a}{2^{k-2}}$, $\frac{b}{2^{k-2}}$, $\frac{c}{2^{k-2}}$
and $\frac{d}{2^{k-2}}$ are integers. Following the conclusions in
Case $k=2$, it can be shown that $\frac{a}{2^{k-2}}$,
$\frac{b}{2^{k-2}}$, $\frac{c}{2^{k-2}}$ and $\frac{d}{2^{k-2}}$ are
even integers, i.e., $2^{k-1}|a$, $2^{k-1}|b$, $2^{k-1}|c$ and
$2^{k-1}|d$.

Then,
$4|(\frac{a}{2^{k-1}})^2+(\frac{b}{2^{k-1}})^2+(\frac{c}{2^{k-1}})^2+(\frac{d}{2^{k-1}})^2$
where $\frac{a}{2^{k-1}}$, $\frac{b}{2^{k-1}}$, $\frac{c}{2^{k-1}}$
and $\frac{d}{2^{k-1}}$ are integers. Applying the conclusions in
Case $k=1$, we have $2|\frac{a}{2^{k-1}}$, $2|\frac{b}{2^{k-1}}$, $
2|\frac{c}{2^{k-1}}$, $2|\frac{d}{2^{k-1}}$ at the same time, or
$2\nmid\frac{a}{2^{k-1}}$, $2\nmid\frac{b}{2^{k-1}}$, $
2\nmid\frac{c}{2^{k-1}}$, $2\nmid\frac{d}{2^{k-1}}$ at the same
time. Hence, $2^{k}|a$, $2^{k}|b$, $2^{k}|c$, $2^{k}|d$ at the same
time, or $2^{k}\nmid a$, $2^{k}\nmid b$, $2^{k}\nmid c$, $2^{k}\nmid
d$ at the same time.

Therefore, Lemma \ref{lemma1_conclusion} is proved.
\end{IEEEproof}

\begin{lemma}\label{lemma2_conclusion}
\emph{For integers $a,b,c,d,e,f,g\emph{\emph{ and }}h$, if
$a^2+b^2+c^2+d^2=e^2+f^2+g^2+h^2$ and $2^k|a^2+b^2+c^2+d^2$ where
$k$ is an integer, then}
\begin{equation*}
2^{k}|ae+bf+cg+dh+af-be+ch-dg.
\end{equation*}
\end{lemma}
\begin{IEEEproof}
Since $a^2+b^2+c^2+d^2=e^2+f^2+g^2+h^2$ and
$2^k|a^2+b^2+c^2+d^2$, we have
$2^{2k}|(a^2+b^2+c^2+d^2)(e^2+f^2+g^2+h^2)$. Let $t_1=ae+bf+cg+dh$,
$t_2=af-be+ch-dg$, $t_3=ag-bh-ce+df$ and $t_4=ah+bg-cf-de$, we have
\begin{equation*}
t^2_1+t^2_2+t^2_3+t^2_4=(a^2+b^2+c^2+d^2)(e^2+f^2+g^2+h^2).
\end{equation*}
Hence, $2^{2k}|t^2_1+t^2_2+t^2_3+t^2_4$.

From Lemma \ref{lemma1_conclusion}, we have $2^{k-1}|t_1$,
$2^{k-1}|t_2$. And $2^{k}|t_1$, $2^{k}|t_2$ at the same time, or
$2^{k}\nmid t_1$, $2^{k}\nmid t_2$ at the same time.

1) When $2^{k}|t_1$, $2^{k}|t_2$ at the same time,
$2^{k}|t_1+t_2$;

2) When $2^{k}\nmid t_1$, $2^{k}\nmid t_2$ at the same time,
then $t_1=2^{k-1}m_1$ and $t_2=2^{k-1}m_2$ where $m_1$ and $m_2$ are
odd integers. Thus, $(t_1+t_2)\emph{\emph{ mod }}2^k=[2^{k-1}(m_1+m_2)]\emph{\emph{ mod }}2^k=2^{k-1}[(m_1+m_2)\emph{\emph{ mod }}2]=0$, i.e., $2^{k}|t_1+t_2$.

Combining the two conclusion, $2^{k}|t_1+t_2$, i.e., $2^{k}|ae+bf+cg+dh+af-be+ch-dg$ holds.
\end{IEEEproof}

\vspace{0.05in}
In the following, we prove Lemma \ref{lemma3_conclusion} based on Lemma \ref{lemma2_conclusion}.

In \textbf{Case I}, $| \Delta s_1|^2+| \Delta s_2|^2=| \Delta s_3|^2+| \Delta s_4|^2$. For integer-coordinate signal
constellations, the difference symbols can be denoted as
\begin{equation*}
\begin{split}
\Delta s_1=&a+bj,\\
\Delta s_2=&c+dj,\\
\Delta s_3=&e+fj,\\
\Delta s_4=&g+hj
\end{split}
\end{equation*}
where $a,b,c,d,e,f,g\emph{\emph{ and }}h$ are integers and will not be zeros at the same time, i.e., $| \Delta s_1|^2+| \Delta s_2|^2=a^2+b^2+c^2+d^2=| \Delta s_3|^2+| \Delta s_4|^2=e^2+f^2+g^2+h^2\neq 0$. Hence, we have
\begin{subequations}
\begin{align*}
\tilde d_1
=&(| \Delta s_1|^2+| \Delta s_2|^2)(u-v)\\
=&(a^2+b^2+c^2+d^2)(u-v),\\
\tilde d_2
=&(\Delta s_1\Delta s^*_3+\Delta s_2\Delta s^*_4)^R-(\Delta s_1\Delta s^*_3+\Delta s_2\Delta s^*_4)^I\\
=&ac+bf+cg+dh+af-be+ch-dg.
\end{align*}
\end{subequations}

Let $a^2+b^2+c^2+d^2$ be expressed as $a^2+b^2+c^2+d^2=2^km$ where $k$ is a non-negative integer and $m$ is an odd integer. Following Lemma \ref{lemma2_conclusion}, it can be shown that $2^k|ac+bf+cg+dh+af-be+ch-dg$, i.e., $2^k|\tilde d_2$. Hence, $\tilde d_2=2^kn$ where $n$ is an integer and we have
\begin{equation*}
\begin{split}
 &\underset{\Delta s_1\emph{\emph{ to }}\Delta s_4}{\min}| \tilde d_1-\tilde d_2|\\
=&\underset{m,n}{\min}| 2^km(u-v)-2^kn|\\
=&\underset{m,n}{\min}(2^k|m(u-v)-n|)\\
=&\underset{m,n}{\min}| m(u-v)-n|.
\end{split}
\end{equation*}

Since $n$ is an integer decided by $\tilde d_2$ and $m$ is an odd integer, we have $\underset{m,n}{\min}| m(u-v)-n|\leq 1/2$. The equality holds if and only if $u-v=\pm 1/2$. Since the coding gain of $\textbf{X}$ in \textbf{Case I} is $\underset{\Delta \textbf{X}}{\min}[det(\Delta \textbf{X}\cdot \Delta \textbf{X}^H)]=\underset{\Delta s_1\emph{\emph{ to }}\Delta s_4}{\min}(2|\tilde d_1-\tilde d_2|^2)$, it is easy to see that $\underset{\Delta \textbf{X}}{\min}[det(\Delta \textbf{X}\cdot \Delta \textbf{X}^H)]\leq1/2$ and the equality holds if and only if $u-v=\pm 1/2$.

Hence, Lemma \ref{lemma3_conclusion} is proved.


\begin{thebibliography}{100}

\bibitem{Zheng}L. Zheng and D. Tse, \textquoteleft \textquoteleft Diversity and multiplexing: A fundamental tradeoff in multiple antenna channels,\textquoteright\textquoteright~\emph{IEEE Trans. Inf. Theory}, vol. 49, no. 5, pp. 1073-1096, May 2003.

\bibitem{Belfiore}J.-C. Belfiore, G. Rekaya, and E. Viterbo, \textquoteleft \textquoteleft The Golden code: A 2x2 full-rate space-time code with non-vanishing determinants,\textquoteright\textquoteright~\emph{IEEE Trans. Inf. Theory}, vol. 51, pp. 1432-1436, Apr. 2005.

\bibitem{Dayal}P. Dayal and M. K. Varanasi, \textquoteleft \textquoteleft An optimal two transmit antenna space-time code and its stacked extensions,\textquoteright\textquoteright~\emph{IEEE Trans. Inf. Theory}, vol. 51, no. 12, pp. 4348-4355, Dec. 2005.

\bibitem{Paredes}J. Paredes, A. B. Gershman, and M. G. Alkhanari, \textquoteleft \textquoteleft A new full-rate full-diversity space-time block code with nonvanishing determinants and simplified maximum-likelihood decoding,\textquoteright\textquoteright~\emph{IEEE Trans. Signal Process.}, vol. 56, pp. 2461-2469, June 2008.

\bibitem{Sezginer}S. Sezginer, H. Sari and E. Biglieri, \textquoteleft \textquoteleft On high-rate full-diversity 2x2 space-time codes with low-complexity optimum detection,\textquoteright\textquoteright~\emph{IEEE Trans. Commum.}, vol. 57, no. 5, pp. 1532-1541, May 2009.

\bibitem{Rabiei}P. Rabiei, N. Al-Dhahir and R. Calderbank, \textquoteleft \textquoteleft New rate-2 STBC design for 2 TX with reduced-complexity maximum likelihood decoding,\textquoteright\textquoteright~\emph{IEEE Trans. Wireless Commun.}, vol. 8, no. 4, pp. 1803-1813, Apr. 2009.

\bibitem{Sezginer2}S. Sezginer and H. Sari, \textquoteleft \textquoteleft A full-rate full-diversity 2x2 space-time code for mobile WiMAX systems,\textquoteright\textquoteright~in~\emph{Proc. ICSPC\textquoteright07}, Dubai, UAE, Nov. 2007.

\bibitem{Biglieri}E. Biglieri, Y. Hong and E. Viterbo, \textquoteleft \textquoteleft On fast-decodable space-time block codes ,\textquoteright\textquoteright~\emph{IEEE Trans. Inf. Theory}, vol. 55, no. 2, pp. 524-530, Feb. 2009.

\bibitem{Tavildar}S. Tavildar and P. Viswanath, \textquoteleft \textquoteleft Approximately universal codes over slow fading channels,\textquoteright\textquoteright~\emph{IEEE Trans. Inf. Theory}, vol. 52, pp. 3233-3258, Jul. 2006.

\bibitem{Tarokh}V. Tarokh, N. Seshadri, and A. R. Calderbank, \textquoteleft \textquoteleft Space-time codes for high data rate wireless communication: performance criterion and code construction,\textquoteright\textquoteright~\emph{IEEE Trans. Inf. Theory}, vol. 44, pp. 744-765, Mar. 1998.

\bibitem{Beesack}P. R. Beesack, \textquoteleft \textquoteleft On the rank of a matrix,\textquoteright\textquoteright~\emph{Mathematics Magazine}, vol. 35, no. 2, pp. 73-77, Mar. 1962.

\bibitem{xia_rotation}W. Su and X.-G. Xia, \textquoteleft \textquoteleft Signal constellations for quasi-orthogonal space-time block codes with full diversity,\textquoteright\textquoteright~\emph{IEEE Trans. Inf. Theory}, vol. 50, no. 10, pp. 2331-2347, Oct. 2004.

\bibitem{Miller}S. L. Miller and R. J. O\textquoteright Dea, \textquoteleft \textquoteleft Peak power and bandwidth efficient linear modulation,\textquoteright\textquoteright~\emph{IEEE Trans. Commum.}, vol. 46, no. 12, pp. 1639-1648, Dec. 1998.

\bibitem{Proakis}J. G. Proakis, \emph{Digital Communications}, 4th ed. New York: McGraw-Hill, 2001.

\bibitem{dvb-s2}ETSI TR 102 376 V1.1.1 (2005-02), Digital Video Broadcasting (DVB)-User Guidelines for the second generation system for Broadcasting, Interactive Services, News Gathering and other broadband satellite applications (DVB-S2), 2005.

\end{thebibliography}
\end{document}